%% file: main.tex
\documentclass[nofootinbib,superscriptaddress,a4paper,twocolumn, floatfix,aps]{revtex4-2}
\usepackage{subfigure}
\usepackage{subfloat}
\usepackage{chemfig}
\usepackage[english]{babel}
\usepackage[utf8]{inputenc}
\usepackage{blindtext}
\usepackage{amsthm}
\usepackage{mathtools}
\usepackage{braket}
\usepackage{xcolor}
\usepackage{graphicx}
\usepackage[left=23mm,right=13mm,top=35mm,columnsep=15pt]{geometry} 
\usepackage[font=small]{caption}
\usepackage{algpseudocode}
\usepackage{bm}
\usepackage{adjustbox}
\usepackage{placeins}
\usepackage[T1]{fontenc}
\usepackage{lipsum}
\usepackage{float}
\usepackage{csquotes}
\usepackage[pdftex, pdftitle={Article}, pdfauthor={Author}]{hyperref}
\usepackage{xcolor}
\usepackage{dblfloatfix}
\usepackage{quantikz}
\hypersetup{
    colorlinks,
    linkcolor={red!50!black},
    citecolor={blue!50!black},
    urlcolor={blue!80!black}
}
\usepackage{booktabs}
\usepackage{multirow}
\usepackage{listings}
\usepackage{xcolor}

\usepackage{caption}
\usepackage{subcaption}
\definecolor{codegreen}{rgb}{0,0.6,0}
\definecolor{codegray}{rgb}{0.5,0.5,0.5}
\definecolor{codepurple}{rgb}{0.58,0,0.82}
\definecolor{tqblue}{HTML}{08293d}
\definecolor{backcolour}{HTML}{fefdf5}

\definecolor{unia}{HTML}{3C2673}
\definecolor{fai}{HTML}{009b5d}
\definecolor{fai2}{HTML}{66c39e}
\definecolor{fai3}{HTML}{b3e1ce}
\definecolor{unia2}{HTML}{8c358b}
\definecolor{unia3}{HTML}{dbbfdd}
\definecolor{unia4}{HTML}{eddeee}

\lstdefinestyle{mystyle}{
    backgroundcolor=\color{backcolour},   
    commentstyle=\color{codegreen},
    keywordstyle=\color{magenta},
    numberstyle=\tiny\color{codegray},
    stringstyle=\color{codepurple},
    basicstyle=\ttfamily\footnotesize\color{tqblue},
    breakatwhitespace=false,         
    breaklines=true,
    postbreak=\mbox{\textcolor{magenta}{$\hookrightarrow$}\space},                 
    captionpos=b,                    
    keepspaces=true,                 
    numbers=left,                    
    numbersep=5pt,                  
    showspaces=false,                
    showstringspaces=false,
    showtabs=false,                  
    tabsize=2
}

\newcommand{\rev}[1]{\textcolor{black}{{#1}}}

\begin{document}

\title{Consistent Initial States with Constant Circuit Depth for\\ Quantum Computational Chemistry}

\author{Lily Barta}
\affiliation{{Institute for Computer Science, University of Augsburg, Germany }}

\author{Jakob~S.~Kottmann}
\email[E-mail:]{jakob.kottmann@uni-a.de}
\affiliation{{Institute for Computer Science, University of Augsburg, Germany }}
\affiliation{{Center for Advanced Analytics and Predictive Sciences, University of Augsburg, Germany }}

\date{\today}

\begin{abstract}
Variational quantum eigensolvers have been extensively studied, yet there are still no methods that offer black-box applicability with consistent performance.
Separable pair approximations promise to be candidates for such methods: they compile to shallow constant-depth quantum circuits with linear gate count and parameter dependence and circumvent most bottlenecks of variational quantum algorithms through their classical simulability. At the same time, they seamlessly integrate into prominent more general circuit designs and subspace strategies. So far, their capability as a consistent method has only been indicated and demonstrations have been restricted to manually designed model systems.  
In this work, we extensively evaluate the consistency of SPA states for hydrogen chains, alkanes, and small molecules within an orbital-optimized VQE framework.
Our benchmarks demonstrate consistent approximations with classical complexity comparable to Hartree-Fock.
\rev{While SPA can serve as a useful low-cost standalone approximation, its restriction to the seniority-zero subspace makes it best suited as a scalable, chemically consistent initial state for more expressive correlation treatments when high quantitative accuracy is required.}
Our open-source implementation within the \textsc{tequila} framework allows convenient use of the algorithms as a standalone method or as a subpart of more extensive procedures.
\end{abstract}

\maketitle

\input{textbody}

\bibliography{main.bib}
\clearpage
\onecolumngrid
\appendix
\input{appendix}
\end{document}

%% file: textbody.tex
\section{Introduction}

One of the central goals of quantum chemistry is the accurate determination of the ground-state properties of molecular systems, which requires solving for the lowest eigenvalue and corresponding eigenstate of the electronic Hamiltonian.
The exact classical solution, within a fixed basis, obtainable via full configuration interaction (FCI), requires a w.r.t system size exponential number of parameters to represent the corresponding state, making it feasible only for small systems. 

Quantum computing has emerged as a promising alternative for electronic structure problems \cite{cao2019quantum,mcardle2020quantum}, offering the possibility to efficiently simulate many-electron wavefunctions. 
In particular, quantum phase estimation (QPE) \cite{aspuru2005simulated} provides a route to efficiently obtain exact eigenvalues and eigenstates from approximate trial states. While initial resource estimates have been cautious about the prospect of routinely applying such computations in feasible time~\cite{reiher2017elucidating}, algorithmic developments over the last decade (see Fig.~2 of Ref.~\cite{babbush2025grandchallengequantumapplications} for an overview) have reduced the estimated runtime of QPE significantly.~\cite{caesura2025faster, gratsea2025achieving}
Despite significantly more optimistic estimates, the computational cost of QPE remains high. Moreover, the algorithm relies on the availability of an initial state with sufficient overlap with the ground state, and therefore requires approximate wavefunctions that can be prepared efficiently while retaining high overlap. Despite potential no-go scenarios with exponentially vanishing overlaps~\cite{Lee2023, louvet2026feasibility} (so-called ``orthogonal catastrophe''~\cite{anderson1967orthogonal}), the high-cost of the algorithm make the overlap of the trial function also a critical factor for feasibility in general. 

The Variational Quantum Eigensolver (VQE) 
\cite{peruzzo2014variational,mcclean2016theory,bharti2022noisy} has originally been proposed as a cheaper alternative to QPE for noisy intermediate-scale quantum (NISQ) devices but already with the prospect of becoming a preceding element that takes care of the trial state in the long run.
VQE is a hybrid quantum–classical variational algorithm in which a parametrized wavefunction is prepared on a quantum device and its energy is evaluated as the expectation value of the electronic Hamiltonian. 
The parameters defining the state are iteratively optimized on a classical computer to minimize this energy, yielding an approximation to the ground state. 

However, practical implementations of VQE on current NISQ hardware remain constrained by the limited number of qubits, gate noise, and restricted circuit depth. While hardware noise is expected to decrease as fault-tolerant quantum computers are developed, other challenges are likely to persist even in the long term. In particular, the large measurement overhead~\cite{gonthier2020identifying, patel2025measurement} required for accurate energy estimation and the occurrence of barren plateaus in the optimization landscape \cite{mcclean2018barren} can significantly hinder the scalability and trainability of variational algorithms.
Two main challenges are therefore the design of expressive yet compact circuit designs, and the efficient implementation and optimization of the corresponding circuit parameters.
Widely used approaches such as the unitary coupled cluster (UCC), in particular UCC with singles and doubles (UCCSD), provide a, in principle, systematic and chemically motivated framework  \cite{romero2019strategies, anand2022quantum}, but typically require a large number of variational parameters and deep circuits, limiting their applicability on near-term devices and complicating their scalability to larger systems. Besides these limitations there is also no convincing indicator with respect to convergence and expected cost of the optimization.

To address these limitations, reduced-complexity and hardware-efficient designs have been proposed. 
Among these, paired-electron approaches restrict the wavefunction to the seniority-zero subspace, thereby reducing the effective Hilbert space while retaining the ability to capture important correlation effects.
To circumvent the costly optimization of the quantum circuit on an actual quantum processor, several works proposed to initialize or approximate quantum circuits with solutions from their non-unitary classically tractable variants (e.g. UCCSD with CCSD/MP2~\cite{romero2019strategies}, oo-pCCD with UpCCD~\cite{krompiec2026simple}, or perturbative approximations~\cite{patel2026quantum}).
To mitigate circuit depth, several circuit design ideas based on adaptive schemes~\cite{grimsley2019adaptive, ryabinkin2018qubit, lang2020iterative}, subspace expansions ~\cite{stair2020multireference, kottmann2024Quantum, park2026canoe} or valence bond heuristics~\cite{burton2022exact, kottmann2023Molecular, ghasempouri2023modular, burton2024Accurate, marti-dafcik2025spin} have been developed by various groups. From a bird's-eye view, all these methods can be seen as extensions to separable circuit design (originating from~\cite{kottmann2022Optimized}) used in this work. 
The separable pair ansatz (SPA) \cite{kottmann2022Optimized}, further exploits this idea by approximating the many-electron wavefunction as a product of independent two-electron wavefunctions. 
This construction leads to a significant simplification of the quantum circuit, with a constant circuit depth and a number of variational parameters that scales linearly with system size. 
At the same time, this separability introduces an approximation through the neglect of inter-pair correlations. 
Importantly, the resulting ansatz has a highly constrained structure that enables efficient classical simulation, and the corresponding quantum circuits are extremely shallow. 
As a consequence, several of the typical challenges associated with general VQE ansätze, such as deep circuit requirements, measurement overheads, and noise accumulation, are avoided in this setting.\\

Despite the various approaches explored in the literature, there is still no consistent VQE method that can be applied with the same convenience as standard approaches from classical quantum chemistry. On the contrary, VQE methods are infamous for producing inconsistent results already within the same molecular instance as well as requiring extensive manual tuning. Circuits based on the separable pair approximation, showed potential for the definition of such a method, as they could be employed as reliable models in various applications~\cite{weber2022Reliability, schleich2021improving, schleich2023partitioning, bincoletto2026transfer, delarcosantos2025Hybrid, gil2025sharc}, so far this however either required the use of pair-natural orbitals~\cite{kottmann2020reducing} or restricted usage to model systems paired with manual orbital guesses~\cite{kottmann2023Molecular}.   

In this work, we give a unified description of SPA circuits and benchmark them across representative chemical systems. We evaluate their performance in terms of ground-state energies, wavefunction fidelities, and computational cost, and analyze their scaling with system size.
In general, benchmarking variational quantum algorithms is challenging due to the strong dependence on ansatz design and the need for instance-specific parameter tuning, which limits the availability of systematic comparisons in the literature. Notable exceptions include adaptive approaches such as ADAPT-VQE~\cite{grimsley2019adaptive} and its variants, which are however computationally expensive and notoriously converge slowly for strongly correlated systems.
In this context, the goal of our study is not only to assess accuracy, but also to illustrate the stability and consistency of the SPA across different systems.\\

In summary we provide
\begin{itemize}
    \item A unified framework for the separable pair approximation (SPA) with detailed analysis on the expected cost (cubic runtime, constant circuit depth with system size, linear circuit depth with basis size)
    \item A detailed description on purely classical determination of the circuit parameters
    \item Systematic benchmarks on the consistency and computational \rev{efficiency} of the method
    \item \rev{Open-source implementations that provide a black-box workflow for hydrogen chains and a chemically motivated initialization procedure for general molecules.}
\end{itemize}
As the SPA method integrates seamlessly into several other circuit designs (see~\cite{kottmann2022Optimized, kottmann2023Molecular, kottmann2024Quantum}), we consider it useful as a baseline for variational quantum eigensolvers, an affordable initial state for QPE, VQE or subspace expansions, that can replace the usual Hartree-Fock (HF) initialization, and a useful testbed for machine learning frameworks (see for example~\cite{bincoletto2026transfer}).

\section{Methodology}
\label{sec:methodo}

\subsection{Separable Pair Ansatz}
\label{sec:spa}

In this work, we employ the SPA, originally introduced in \cite{kottmann2022Optimized}, in which an $N$-electron wavefunction is approximated as a product of $N/2$ two-electron wavefunctions. 
Assuming $N$ is even (see the appendix of Ref.~\cite{kottmann2023Molecular} for an illustration on odd numbers), the total SPA wavefunction can be expressed as
\begin{equation}
    \ket{\Psi_\text{SPA}} = \bigotimes_{k=1}^{N/2} \ket{\psi_{k}},
\end{equation}
where each $\ket{\psi_{k}}$ describes a correlated electron pair, restricted to its own set of spatial orbitals $S_k$. 
The full orbital space is therefore partitioned into disjoint subsets, such that orbitals belong to exactly one electron pair.
The choice of electron pairs, together with orbital optimization, is crucial for the performance of the SPA ansatz and is discussed further in Section \ref{sec:orb-opt}.

To prepare the SPA wavefunction, we apply a sequence of unitary operators to a closed-shell reference state \[\ket{\Psi_0} = \bigotimes_{k=1}^{N/2} \left(\ket{11} \otimes \ket{0}^{\otimes2|S_k|-2}\right),\]
\begin{equation}
    \ket{\Psi_\text{SPA}} = \prod_{k=1}^{N/2} U_k(\theta_k) \ket{\Psi_0}, \label{eq:ref-state}
\end{equation}
where each $U_k$ acts exclusively within its associated orbital subset $S_k$.
These unitary operators generate paired-electron excitations and are defined as
\begin{equation}
    U_k(\theta_k) = \prod_{l=1}^{|S_k|-1} e^{\frac{\theta_k^l}{2}
    (a_{l_\uparrow}^\dagger a_{(l+1)_\uparrow}a_{l_\downarrow}^\dagger a_{(l+1)_\downarrow}-h.c.)
    }
\end{equation}
where $\boldsymbol{\theta}_k = \{\theta_k^l\}$ are the variational parameter associated with the $k$-th electron pair.
The parameters $\boldsymbol{\theta} = \{\theta_k\}$ are optimized with VQE by minimizing the electronic Hamiltonian expectation value,  
\begin{equation}
    E(\boldsymbol{\theta}) = \min_{\boldsymbol{\theta}}  
\bra{\Psi_\text{SPA}(\boldsymbol{\theta})} \hat{H} \ket{\Psi_\text{SPA}(\boldsymbol{\theta})}.
\end{equation}

The structure of these operators $U_k$ is analogous to those used in the paired-electron Unitary Coupled Cluster with Doubles (pUCCD) ansatz \cite{elfving2021simulating}.
\rev{SPA can be viewed as a restricted, separable form of pUCCD: while the latter allows pair excitations across the entire orbital space, SPA partitions the orbital space into disjoint subsets $S_k$ and restricts pair excitations to occur only within each subset. 
Consequently, correlations between different electron pairs are excluded.
The pUCCD ansatz can thus be viewed as an extension of SPA in which the restriction to intra-subset pair excitations is removed.}

\subsection{Hard-Core Boson Model}

Because we start from a closed-shell reference and apply only paired-electron excitations, the ansatz remains entirely within the seniority-zero subspace. 
That is, all configurations generated consist exclusively of empty or doubly occupied spatial orbitals.

This restriction allows us to reformulate the problem in terms of electron-pair creation and annihilation operators rather than individual fermionic operators. 
Introducing the pair operators
\begin{equation}
    b_{l}^\dagger = a_{l_\uparrow}^\dagger a_{l_\downarrow}^\dagger,
    \quad b_{l} = a_{l_\uparrow} a_{l_\downarrow}
\end{equation}
that create and annihilate quasi particles consisting of spin-paired electrons, often referred to as hard-core Bosons.
In this encoding, the unitary operators from above become
\begin{equation}
    U_k(\theta_k) = \prod_{l=1}^{|S_k|-1} e^{\frac{\theta_k^l}{2}(b_{l}^\dagger b_{(l+1)}-h.c.)}.
\end{equation}

This reduced description is referred to as the hard-core boson (HCB) model and is the standard formulation for SPA and pUCCD~\cite{elfving2021simulating} approaches.

\subsection{Qubit Mapping} \label{qubit-mapping}

For a qubit implementation, we need to map the HCB operators to Pauli operators. 
This can be achieved using an analogy of the Jordan–Wigner transformation, where pair creation and annihilation operators are mapped to qubit raising and lowering operators,
\begin{equation}
    b_{l}^\dagger \longrightarrow \sigma_l^-,
    \quad b_{l} \longrightarrow \sigma_l^+
\end{equation}
with
\begin{equation}
    \sigma_l^+ = \frac{1}{2} (X(l) + iY(l)),
    \quad \sigma_l^- = \frac{1}{2} (X(l) - iY(l)).
\end{equation}
Since the HCB operators commute, no anti-symmetry is necessary here.
In this encoding, each spatial orbital is represented by a single qubit, which is in the state $\ket{1}$ if the orbital is doubly occupied and $\ket{0}$ if it is empty. 
Because the ansatz is restricted to the seniority-zero subspace, this mapping reduces the number of qubits by a factor of two compared to the standard fermion-to-qubit encoding.

The pair-excitation generator between two spatial orbitals $k$ and $l$ can then be written in terms of Pauli operators
\begin{equation}  
b_k^\dagger b_{l} - b_{l}^\dagger b_k  
\;\longrightarrow\;  
\frac{i}{2}\left(X(k)Y(l) - Y(k)X(l)\right),  
\end{equation}
and the corresponding unitary operator acting on qubits becomes  
\begin{equation}
    U_{l,l+1}(\theta_k^l) = e^{i\frac{\theta_k^l}{4}\left(X(l)Y(l+1) - Y(l)X(l+1)\right)}
\end{equation}
This operator acts nontrivially only on the two-qubit subspace spanned by qubits $l$ and $(l+1)$, and is implicitly tensored with identity operators on all other qubits
\begin{equation}
    U_{l,l+1}(\theta_k^l) \equiv I \otimes \dots \otimes U_{l,l+1}(\theta_k^l) \otimes \dots \otimes I.
\end{equation}
In the computational basis $\{\ket{00}, \ket{01}, \ket{10}, \ket{11}\}$, its matrix representation is
\begin{equation}  
U_{l,l+1}(\theta_k^l) =
\begin{pmatrix}  
    1       & 0                   & 0                  & 0   \\
    0       & \cos(\theta_k^l/2)  & \sin(\theta_k^l/2) & 0   \\  
    0       & -\sin(\theta_k^l/2) & \cos(\theta_k^l/2) & 0   \\
    0       & 0                   & 0                  & 1   
\end{pmatrix}  
\end{equation}  
Thus the operator acts as a Givens rotation in the singly occupied hardcore-bosonic subspace $\{\ket{01}, \ket{10}\}$, while leaving the empty $\ket{00}$ and doubly occupied $\ket{11}$ sectors unchanged.

\subsection{Circuit Construction}

In practice, the unitary operators $U_k$ are implemented using combinations of CNOT and controlled-$R_y$ ($CR_y$) gates, as they suffice to generate the linear combination of unitary basis states that represent a single hardcore-bosonic quasi-particle. 
For a two-electron two-orbital system, the circuit takes the form
\[
    U = \adjustbox{valign=c}{\includegraphics[width=0.6\columnwidth]{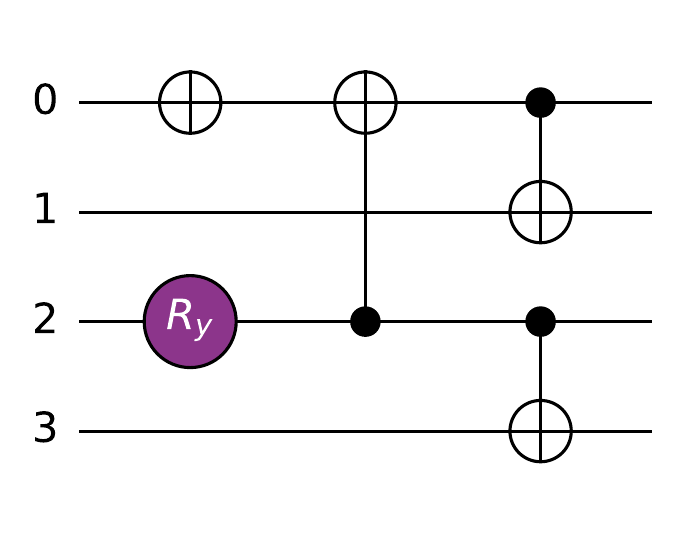}}
\]
The first part of the circuit generates the hardcore-bosonic superposition on qubits 0 and 2
\begin{align}
    \cos(\theta/2)\ket{10} + \sin(\theta/2)\ket{01}.    
\end{align}
\rev{The final two CNOT gates shown in the circuit instead provide an optional embedding into the full fermionic register (see~\cite{kottmann2022Optimized} and ~\cite{kottmann2023Molecular} for more details) 
\begin{align}
    \cos(\theta/2)\ket{1100} + \sin(\theta/2)\ket{0011}
\end{align}
where we have used odd qubits to encode spin-up and even qubits to encode spin-down electrons in a Jordan-Wigner encoding. This embedding is useful when SPA serves as the initial state for a subsequent many-body method that may leave the seniority-zero subspace. For the standard Bravyi-Kitaev encoding with the corresponding spin-orbital ordering, the full register can instead be constructed by adding one qubit initialized in $\ket{0}$ for each spatial orbital, without the final CNOT layer.}
Note that  the circuit above does not implement the operation $U_{l,l+1}$ described above, but generates the same parametrized state when applied to the reference state in Eq.~\eqref{eq:ref-state}.

For larger orbital subsets $S_k$, the same construction is extended by adding further layers of $CR_y$ and CNOT gates.
For example, a two-electron four-orbital system can be prepared using
\[
    U = \adjustbox{valign=c}{\includegraphics[width=0.9\columnwidth]{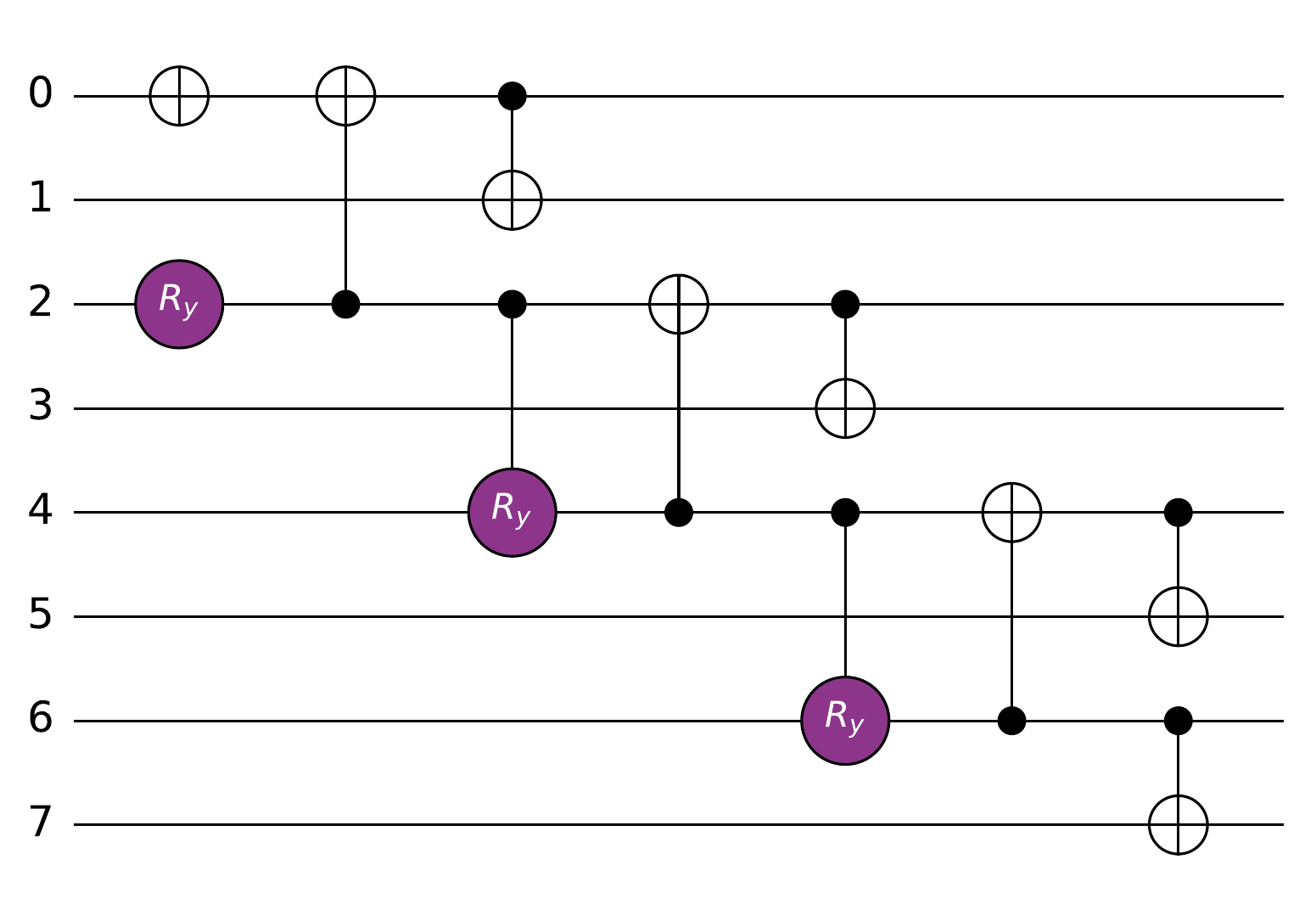}}
\]
where each additional layer introduces excitations into another spatial orbital while preserving the paired-electron structure.

For systems containing multiple electron pairs, separate circuits are constructed for each subset $S_k$ and are executed in parallel. 
As a result, increasing the number of electron pairs does not increase the circuit depth.

\subsection{Chemically Motivated Pair Selection and Orbital Optimization}
\label{sec:orb-opt}

The performance of the SPA depends on the choice of electron pairs and orbital optimization.

The orbital subsets $S_k$, introduced previously in Section \ref{sec:spa}, are defined based on the molecular bonding structure. 
Specifically, we use the connectivity of the molecular chemical graph, which can be interpreted in terms of a Lewis structure, to define the electron pairs. 
For electron pairs corresponding to covalent bonds, this leads to bonding and antibonding-like combinations, while lone pairs are described by localized non-bonding orbitals. 
For simple systems such as hydrogen chains, this can be achieved directly from orthonormalized atomic orbitals by forming symmetric and antisymmetric combinations. 
For more general molecules, we start from HF orbitals and localize them to obtain orbitals corresponding to bonds and lone pairs.

While this construction yields a chemically motivated initial partitioning, the SPA ansatz is not invariant under orbital rotations. 
Consequently, the accuracy of the pair approximation highly depends on the orbital representation, and orbital optimization is required to obtain physically meaningful results. 
As discussed in Ref.~\cite{zhao2023orbital}, neglecting orbital optimization can lead to qualitatively incorrect energy profiles, particularly in bond dissociation regimes.
Here, orbital optimization is performed using a classical optimizer, where the energy is evaluated via SPA-VQE. 
The required reduced density matrices (RDMs) are computed within our framework to exploit the separability of the ansatz and improve computational efficiency, as described in the following section.

\rev{The initialization procedures are specified in Section \ref{sec:algorithm}. Their sensitivity to orbital localization and initial pairing, and the extent to which orbital optimization mitigates this sensitivity, are assessed for butane and butadiene in Appendix \ref{app:orb-loc}.}

\subsection{Efficient Evaluation of Expectation Values}
\label{sec:fast-spa}

As described in Section \ref{qubit-mapping}, operators expressed in second quantization can be mapped to qubit operators using the Jordan–Wigner transformation.
More generally, this mapping transforms fermionic or pair operators into weighted sums of Pauli string operators, i.e., tensor products of the single-qubit operators $I$, $X$, $Y$, and $Z$.
Any such qubit operator $\hat{O}$ can therefore be written as  
\begin{equation}  
\hat{O} = \sum_{i=1}^{|P|} c_i \hat{P}_i,
\label{eq:pauli_sum}
\end{equation}
where each $\hat{P}_i$ is a Pauli string.
In particular, both the electronic Hamiltonian and the operators required for evaluating RDM elements take this form.

The separability of the SPA ansatz can be exploited to evaluate expectation values more efficiently through products of local expectation values.
For this purpose, we introduce the notion of computational clusters. 
These clusters define how the calculation is partitioned and may group one or multiple electron-pair states.
The local state associated with each cluster $C_\alpha$ is given by
\begin{equation}
    \ket{\Psi_\alpha} = \bigotimes_{k \in C_\alpha} \ket{\psi_k}.
\end{equation}

With this definition, each Pauli string factorizes over the computational clusters as
\begin{equation}
    \hat{P}_i = \bigotimes_{\alpha=1}^{N_c} \hat{P}_i^{(\alpha)},
\end{equation}
where $\hat{P}_i^{(\alpha)}$ acts only on the qubits contained in cluster $C_\alpha$.
The expectation value of $\hat{P}_i$ then reduces to
\begin{equation}
\braket{\Psi_\text{SPA}|\hat{P}_i|\Psi_\text{SPA}}
=
\prod_{\alpha=1}^{N_c}
\langle \Psi_\alpha | \hat{P}_i^{(\alpha)} | \Psi_\alpha \rangle.
\end{equation}
Inserting this decomposition into Eq. (\ref{eq:pauli_sum}) yields
\begin{equation}
\braket{\Psi_\text{SPA}|\hat{O}|\Psi_\text{SPA}}
=
\sum_{i=1}^{|P|} c_i \prod_{\alpha=1}^{N_c}
\braket{\Psi_\alpha | \hat{P}_i^{(\alpha)} | \Psi_\alpha}.
\label{eq:fspa}
\end{equation}

The computational cost therefore depends on the chosen cluster size $N_c$, introducing a tradeoff between the number of evaluations and the dimension of the cluster states.
The highest factorization can be achieved with $N_c = N/2$, corresponding to one electron pair per cluster.
To further improve efficiency, local expectation values
$\langle \Psi_\alpha | \hat{P}_i^{(\alpha)} | \Psi_\alpha \rangle$
are computed only once, stored, and reused whenever the same local Pauli string reappears. 

\rev{In this work, expectation values are evaluated
classically by exploiting the separability of SPA according
to Eq.~\ref{eq:fspa}. Energy measurement on quantum hardware is
discussed in the following subsection.}

\subsection{Potential as Hardware Testbed}
\rev{Despite being classically simulable, SPA circuits could be a useful testbed for current and emerging hardware platforms beyond plain initial states. Potential applications could be: benchmarks with varying circuit depth by off-loading orbital rotations from the Hamiltonian to the circuit, guiding states as used in recently developed mitigation schemes~\cite{zou2025quantum}, resource states for sampled subspace diagonalization~\cite{Robledo_Moreno_2025}, or the classically tractable part in recent measurement reduction strategies~\cite{izmaylov2026classicalactivespacehybridquantum}.}\\

\rev{SPA can be implemented either directly in the HCB representation, with one qubit per spatial orbital, or in a full fermionic register, with one qubit per spin orbital. The HCB Hamiltonian contains only diagonal number-operator terms and pair-hopping terms, which can be measured using three settings corresponding to the $Z$, $X$, and $Y$ bases, as for pUCCD~\cite{elfving2021simulating} (see ~\cite{zhao2024enhancing} for an application on ion traps)}
\rev{Three measurement settings also suffice when the SPA state is represented in the full fermionic register. In this case, the uncompressed HCB Hamiltonian is constructed using fermionic operators, following Ref.~\cite{Bincoletto2026Measurement}. Its Pauli representation decomposes into three mutually commuting groups, each measured with an appropriate basis-change circuit. Since SPA is restricted to the seniority-zero subspace, the remaining terms of the electronic Hamiltonian have vanishing expectation values. Consequently, this procedure yields the full electronic energy expectation value of the SPA state without measuring additional residual contributions.}\\

\rev{Note at this point, that these three energy-measurement settings are favorable in the number of circuits to be measured (precicely three) but do not necessarily imply a small total shot cost. We refer to Ref.~\cite{Bincoletto2026Measurement} for a deeper discussion and extensions beyond the hardcore-bosonic regime.}

\subsection{Scaling and Expressivity of SPA}

The separable structure of SPA and the HCB representation lead to favorable scaling for both memory requirements and energy expectation-value evaluation.
To analyze these scalings, we define
\begin{itemize}
    \item $N_p = N/2$: the number of electron pairs,
    \item $n_{pp}$: the number of spatial orbitals per pair,
    \item $N_{orb} = N_p n_{pp}$: the total number of spatial orbitals.
\end{itemize}

Each electron pairfunction $\ket{\psi_k}$ is described within a local $n_{pp}$-qubit space, spanned by basis states containing a single occupied orbital $\{\ket{100\cdots0},\ket{010\cdots0},\ldots,\ket{000\cdots1}\}$ which span a $n_{pp}$-dimensional subspace.
Thus, for single-pair computational clusters, the memory required to represent each local state scales linearly with respect to the assigned orbitals $\mathcal{O}(n_{pp})$.  

The computational cost for evaluating the expectation value in Eq. (\ref{eq:fspa}) therefore scales as
\begin{equation}
    \mathcal{O}(M N_p n_{pp})
    =
    \mathcal{O}(M N_{orb}),
    \label{eq:fspa_scaling}
\end{equation}
where $M$ denotes the number of Pauli strings that represent the observable.
For a conventional fermionic Hamiltonian, the number of Pauli strings generated after fermion-to-qubit mapping grows approximately quartically with the number of spatial orbitals
\begin{equation}
    M_{ferm} \sim \mathcal{O}(N_{orb}^4),
\end{equation}
due to the two-electron interaction term running over four different spatial orbital indices.
Within the HCB approximation (that is exact for this ansatz) this is however reduced to a quadratic number of terms
\begin{equation}
    M_{HCB} \sim \mathcal{O}(N_{orb}^2).
    \label{eq:hcb_ham_scaling}
\end{equation}

Combining Eqs. (\ref{eq:fspa_scaling}) and (\ref{eq:hcb_ham_scaling}) leads to an overall cubic scaling $\mathcal{O}(N_{orb}^3)$ with respect to the total number of spatial orbitals.\\

This cubic scaling is obtained when increasing system size for fixed basis size per pair ($n_{pp} = \mathrm{const.}$) which corresponds to increasing the number of electrons with number of orbitals growing directly proportional to it. This situation for example arises when atom centered basis sets are chosen (e.g. STO-3G or cc-pVDZ). 
The contrary situation: increasing the basis set while keeping the number of electrons constant leads to the same cubic scaling.

Finally, although SPA admits a favorable polynomial computational scaling, it still spans an exponentially large determinant space.
Since the wavefunction contains $N_p$ electron pairs that can each occupy any of the $n_{pp}$ orbitals assigned to its subset, the total number of Slater determinants scales as $n_{pp}^{N_p}$.

\section{Algorithm}\label{sec:algorithm}

We briefly describe the automated initialization procedures implemented using \texttt{Tequila}, its \texttt{sunrise} extension, and the \texttt{quanti-gin} package. At the time the algorithm requires neutral charge singlet states as target (see the appendix of ~\cite{kottmann2023Molecular} for an illustration of open-shell systems). We further consider a minimal basis setting, where for $N_e$ electrons $N_o=N_e$ spatial basis functions are chosen.\\ 

\rev{We will describe the algorithm for two scenarios}
\rev{
\begin{enumerate}
    \item An easy-to-implement version restricted to hydrogenic systems $H_n$ consisting of $n$ hydrogen atoms in minimal bases. 
    \item Typical organic molecules, to demonstrate applicability and potential for chemical systems.
\end{enumerate}
}
\rev{
Due to their conceptual simplicity, hydrogenic systems (case 1) are often used as benchmark and test systems in the context of quantum computations.
}\\

\textit{1. Hydrogenic Models: }
For hydrogenic systems, the initial electron-pair assignment is obtained automatically from the nuclear geometry.
A graph is constructed in which each vertex corresponds to a hydrogen atom and the weight associated with the edge connecting atoms $i$ and $j$ is defined as
\begin{equation}
    w_{ij} = \tanh\left(\frac{\pi}{2}\|\mathbf{r}_i-\mathbf{r}_j\|_2\right).
\end{equation}
where $\left\|\mathbf{r}_i-\mathbf{r}_j\right\|_2$ is the Euclidean norm and the $\tanh$ screening has been proven useful~\cite{bincoletto2026transfer} to take dissociation, after which larger distances should have no further effect, into account.
The pairing is then determined by solving a minimum-weight matching problem which solves as a heuristic model to identify the best Lewis graph of the molecule. This graph selection heuristic selects the accurate Lewis graphs for chemically well interpretable systems (such as linear hydrogen chains and ring systems) and has been performing comparably well on randomized instances (see also the appendix).  
For each selected pair, bonding and antibonding orbitals are constructed from the corresponding atomic orbitals. 
These orbitals define the initial subsets $S_k$ and provide the initial guess for the orbital optimization. 
In matrix form, this initial orbital transformation has a block-diagonal structure composed of $2\times2$ matrices of the form
\begin{equation}
    \begin{pmatrix}
        1 & 1 \\
        1 & -1
    \end{pmatrix}
\end{equation}
for each electron pair.
\rev{The complete procedure is implemented in the quanti-gin package \cite{rabavikt2026quantigin}. Appendix B provides a minimal usage example.}\\

\rev{2. \textit{General molecules: } Here we employ the Chemist’s localized property-optimized orbitals (CLPOs) \cite{nikolaienko2019Localizeda} which we construct with \texttt{sunrise}'s \texttt{JANPA} \cite{nikolaienko2014JANPA} interface. The CLPO procedure optimizes atom-centered hybrid orbitals and their combinations into a Lewis-like description of bonds and lone pairs. The \texttt{sunrise} interface directly supplies the orbital pairing and reorders the orbitals so that each bonding orbital and its corresponding antibonding orbital appear consecutively, defining the orbital subsets \(S_k\). The initial guess for the subsequent orbital optimization is the identity matrix in this localized basis, i.e., no additional orbital rotations are applied initially.}

\rev{Appendix \ref{app:orb-loc} illustrates the advantages by exemplary comparison of this initialization with the more popular Boys and Pipek–Mezey localizations and alternative initial pairings. 
}

\section{Computational Details}
\label{sec:comp-details}

Our data were generated using the \texttt{tequila} Python package \cite{tequila} and its chemistry extension \texttt{sunrise} \cite{sunrise}. 
We used \texttt{Qulacs} \cite{qulacs} as quantum simulation backend, the Jordan-Wigner implementation of \texttt{OpenFermion} \cite{OpenFermion}, \texttt{PySCF} \cite{pyscf1} for molecular integral calculations and \texttt{JANPA} \cite{nikolaienko2014JANPA} for orbital localization. 
Reference FCI, HF and coupled cluster with perturbative triples (CCSD(T)) energies were obtained via Tequila's \texttt{PySCF} interface. The Tequila and Sunrise libraries are available on GitHub. 

\section{Results and Discussion}
\label{sec:results}

In the following, we assess the performance of the SPA using both model and molecular systems.
Linear hydrogen chains are used as benchmark systems for strongly correlated physics, enabling controlled studies of dissociation and system-size scaling. 
We then extend the analysis to molecular systems to evaluate the behaviour of SPA in realistic chemical settings.

\subsection{Hydrogen chains}

Linear hydrogen chains ($\mathrm{H_N}$) serve as strongly correlated model systems and provide a challenging benchmark even for relatively small system sizes. 
In this work, we consider uniformly spaced geometries and vary both system size and interatomic separation. 
By doing so, we assess the stability and consistency, of SPA across different correlation regimes.

For all hydrogen chains, electron pairs are defined between nearest-neighbour atoms along the chain, i.e., pairing orbitals $(i,i+1)$ for even indices $i$. 
All calculations are performed in the minimal STO-3G basis.

\subsubsection{Dissociation of linear H$_6$}

\begin{figure}[h]
    \centering
    \includegraphics[width=\columnwidth]{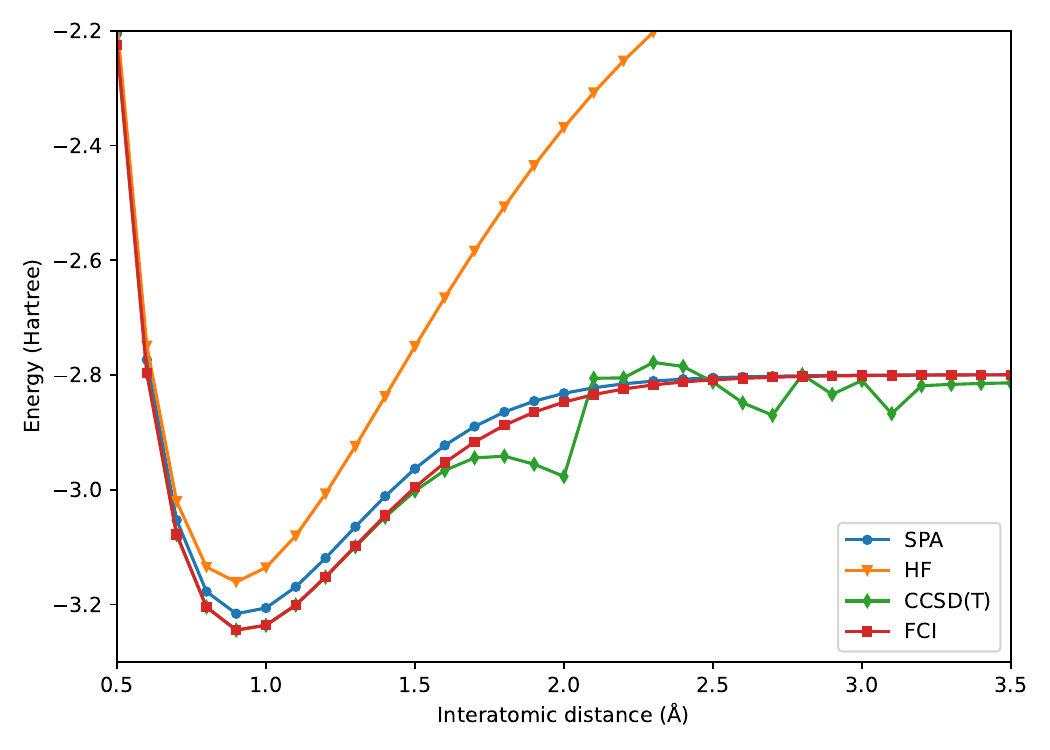}
    \caption{Dissociation energy curves of linear $\mathrm{H_6}$ for SPA, HF, CCSD(T) and FCI.}
    \label{fig:dissoc_h6}
\end{figure}

We first examine the dissociation behaviour of SPA using $\mathrm{H_6}$ as a representative benchmark system. 
Figure~\ref{fig:dissoc_h6} compares the SPA dissociation curve against HF, CCSD(T), and FCI. 

SPA provides a stable and qualitatively accurate description of the dissociation process and remains well-behaved in the stretched-bond regime. 
At the dissociation limit, SPA becomes exact, as the system factorizes into independent atoms and inter-pair correlation vanishes. 
In contrast, both HF and CCSD(T) fail to correctly describe the dissociation process due to their underlying single-reference nature. 
While CCSD(T), often regarded as the gold standard of quantum chemistry, is quantitatively accurate near equilibrium bond lengths, it becomes unstable as the interatomic distance increases and static correlation becomes dominant. 
SPA does not suffer from this breakdown, as its multi-determinant character enables it to capture this static correlation.

\subsubsection{Dissociation across different chain lengths}

\begin{figure}[h]
    \centering
    \includegraphics[width=\columnwidth]{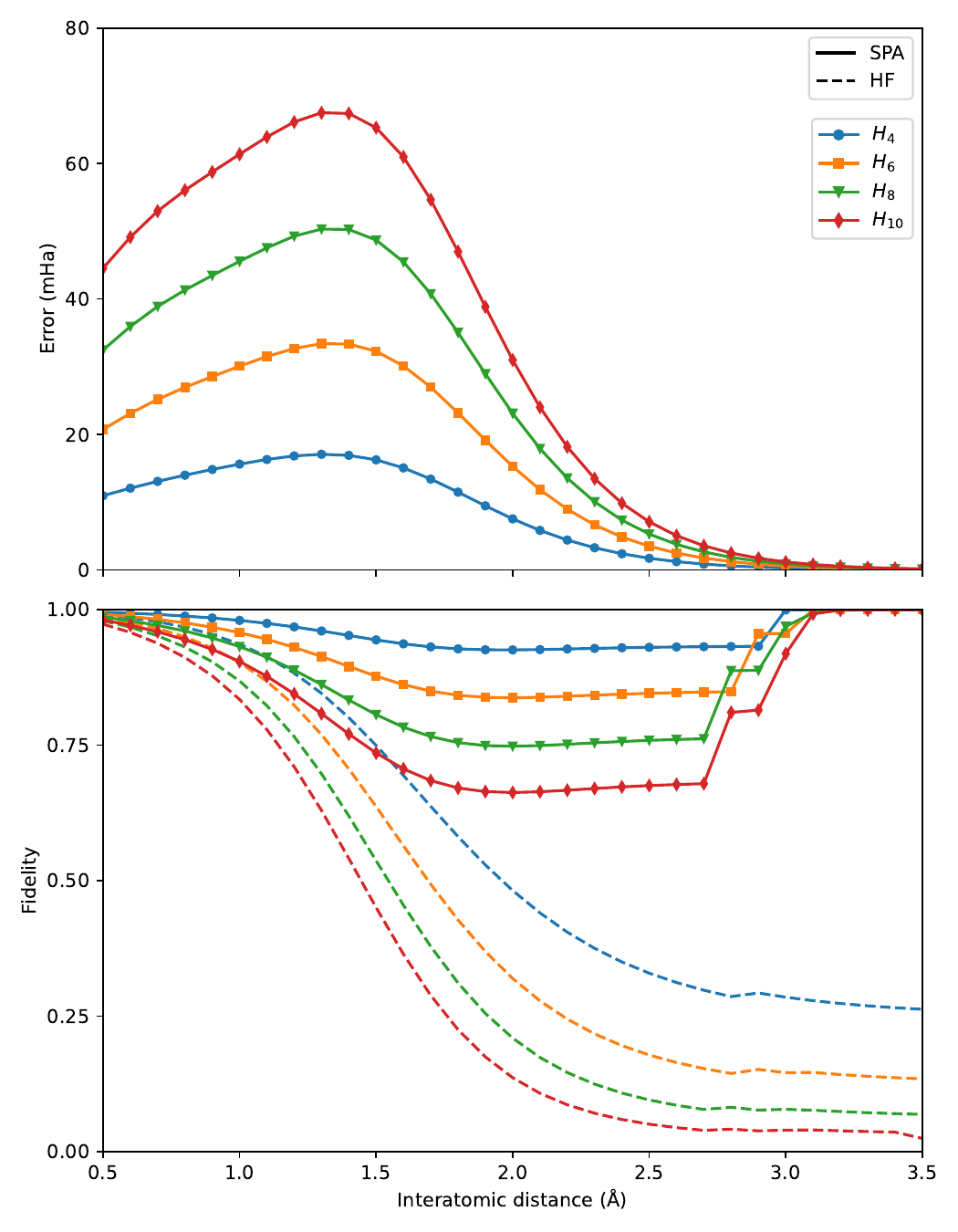}
    \caption{(top) SPA energy error relative to FCI and, (bottom) SPA and HF fidelities $|\braket{\Psi|\Psi_\text{FCI}}|^2$, of linear $\mathrm{H_4}$, $\mathrm{H_6}$, $\mathrm{H_8}$ and $\mathrm{H_{10}}$, along the full dissociation curve. Jumps in the fidelities are caused by convergence problems of the FCI Davidson solver.}
    \label{fig:dissoc_error_fid}
\end{figure}

We next consider hydrogen chains $\mathrm{H_4}$ to $\mathrm{H_{10}}$ to study system-size scaling across the full dissociation coordinate. 
Figure~\ref{fig:dissoc_error_fid} (top) shows the energy error relative to FCI as a function of bond length.

Near equilibrium, the energy error is approximately 15, 29, 43, and 59 mHa for $\mathrm{H_4}$, $\mathrm{H_6}$, $\mathrm{H_8}$, and $\mathrm{H_{10}}$, respectively. 
At intermediate bond distances, the error grows further, reaching maximum values of approximately 17, 33, 50, and 67 mHa around $R \approx 1.4$\AA\, before decreasing and vanishing in the dissociation limit.

The corresponding wavefunction fidelities, $|\braket{\Psi|\Psi_\text{FCI}}|^2$, are shown in Figure~\ref{fig:dissoc_error_fid} (bottom).
Near equilibrium and at short bond lengths, SPA achieves high fidelities. 
As the chains are stretched, the fidelity decreases and reaches a minimum around $R \approx 2$\AA\, reflecting the increasing importance of inter-pair correlation not captured by the ansatz. 
However, in the dissociation limit, the fidelity converges towards one, consistent with the exact factorization into isolated atoms. 
In contrast, HF exhibits systematically lower fidelities and deteriorates rapidly with increasing bond length.

Overall, both the energy error and the fidelity exhibit an approximately linear scaling with system size, consistent with near-extensive behaviour.

\begin{figure}[h]
    \centering
    \includegraphics[height=13.4cm]{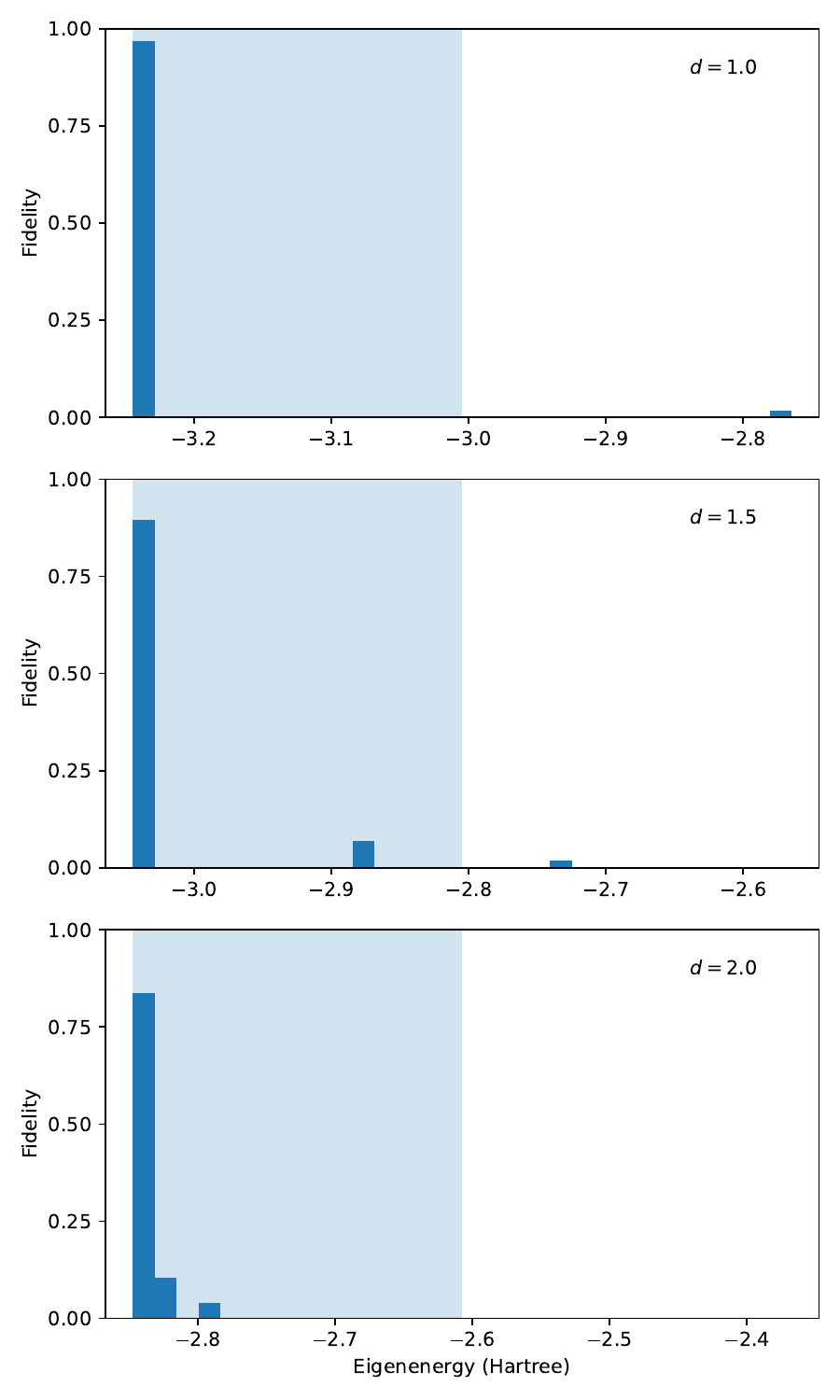}
    \caption{SPA fidelities $|\braket{\Psi_\text{SPA}|\Psi_\text{FCI}}|^2$ for all eigenstates of linear $\mathrm{H_6}$ at interatomic distances 0.9 (top), 1.4 (centre) and 2.0 \AA\ (bottom). The blue shaded region indicates the UV–vis excitation window relative to the ground state.}
    \captionsetup{skip=1pt}
    \label{fig:fid_spectra_h6}
\end{figure}

\subsubsection{Wavefunction analysis via fidelity spectra}

To gain further insight into the origin of the observed energy errors, we analyze the distribution of SPA wavefunction weight across the full eigenstate spectrum of $\mathrm{H_6}$.
Figure \ref{fig:fid_spectra_h6} shows the fidelity spectrum for different interatomic distances: 0.9 (left), 1.4 (centre) and 2.0\AA\ (right).
The blue shaded region indicates the UV–vis (spectroscopic) excitation-energy window measured relative to the ground state. We use this as a stand-in to characterize potentially interesting excited states, that would get samples in a scenario where the state is used as initial state for QPE. 
Most of the SPA wavefunction weight falls within this range.

These results help explain the contrasting trends in energy error and ground state fidelity observed in Figure \ref{fig:dissoc_error_fid}. 
While the fidelity is higher at $R = 1.4$\AA\  than at $R = 2.0$\AA\, the remaining overlap at $R = 1.4$\AA\ is spread over higher-energy excited states.
Since the VQE energy is given by $E = \sum_i |\braket{\Psi^{\text{SPA}}|\Psi_i^{\text{FCI}}} |^2 E_i$, leakage into higher-energy states leads to a larger energy deviation. 
By contrast, at $R = 2.0$\AA\, the residual weight is predominantly distributed among low-lying excited states, resulting in a smaller energy error despite a reduced ground-state fidelity.
In contrast, SPA remains computationally tractable (see next section).

\subsubsection{System-size scaling at fixed interatomic distance}

\begin{figure}[h]
    \centering
    \includegraphics[width=\columnwidth]{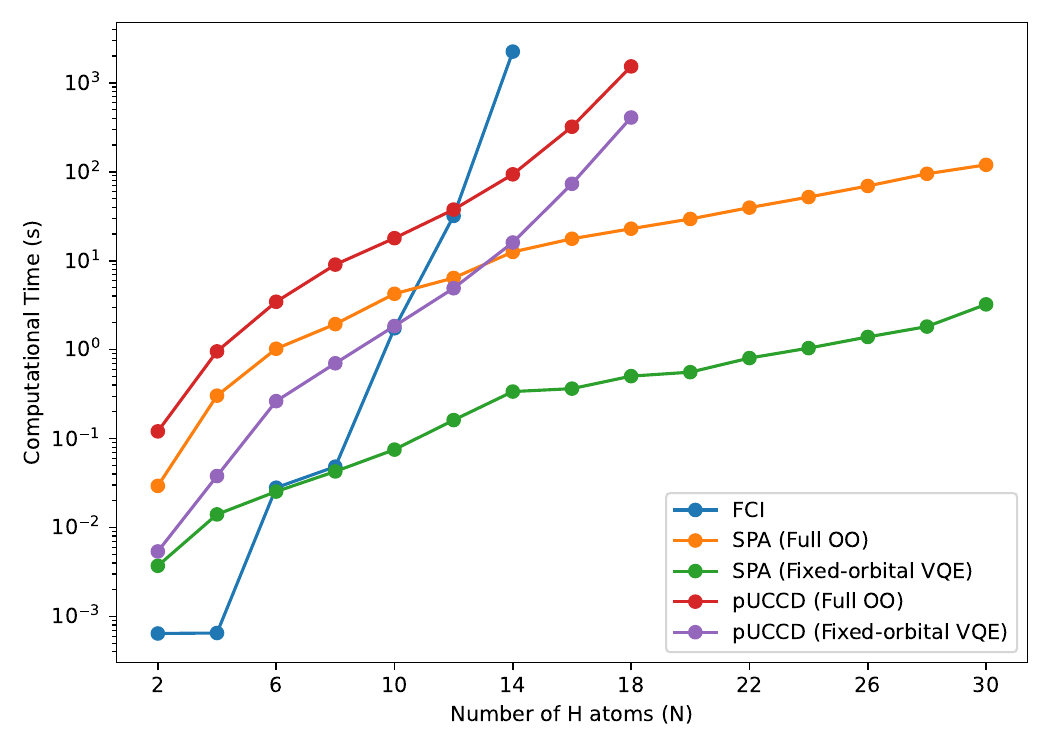}
    \caption{Classical computational cost for linear hydrogen chains at fixed interatomic distance of $R=1.0\,\mathrm{\AA}$. For SPA and pUCCD, "Full OO" denotes the complete orbital-optimized calculation, while "Fixed-orbital VQE" denotes a separate variational-parameter optimization in the final optimized orbital basis.}
    \label{fig:times_vs_n}
\end{figure}

\rev{We next investigate the computational cost and accuracy of SPA as the number of hydrogen atoms is increased at a fixed interatomic distance of $R=1.0~\text{\AA}$. Figure~\ref{fig:times_vs_n} compares the computation times for FCI, SPA, and orbital-optimized pUCCD (oo-pUCCD). Note that, in these calculations, SPA-initialization for the orbitals was required for the oo-pUCCD calculations to converge.
For both SPA and oo-pUCCD, the figure distinguishes the complete orbital-optimized calculation, in which orbital and variational-parameter optimizations are alternated until convergence, from a separate VQE parameter optimization performed in the final optimized orbital basis.
All reported times are classical wall times; in particular, the oo-pUCCD timings correspond to exact classical simulation of the circuit and should not be interpreted as quantum-hardware execution times.}

\rev{At $\mathrm{H_{14}}$, the FCI calculation required 2200~s, whereas the complete orbital-optimized SPA calculation required 12~s. SPA calculations were performed up to the chosen endpoint $\mathrm{H_{30}}$, for which the complete orbital-optimized calculation required 120~s.}

The low computational cost of SPA originates from the separable structure of the ansatz, which leads to a constant circuit depth of 3 and a number of variational parameters that scales linearly with system size, specifically as $N/2$ corresponding to the number of electron pairs. 
As a result, the cost of the VQE step remains small across all system sizes considered, requiring only 3.2~s for $\mathrm{H_{30}}$. 
The dominant computational cost arises from the orbital optimization step, which currently constitutes the primary bottleneck of the method.

\begin{table}[h]
\centering
\setlength{\tabcolsep}{10pt}
\caption{Number of SPA VQE iterations (with pre-optimized orbitals and parameters initialized to zero) and orbital optimization cycles (CASSCF macro-iterations) for H$_6$, H$_{10}$, H$_{30}$.}
\begin{tabular}{c c c}
\hline
System & VQE iterations & Orb. Opt. cycles \\
\hline
H$_6$  & 4 & 5 \\
H$_{10}$ & 4 & 5 \\
H$_{30}$ & 4 & 6 \\
\hline
\end{tabular}
\label{tab:opt_cycles}
\end{table}

Nevertheless, the variational optimization itself remains inexpensive.
Table~\ref{tab:opt_cycles} reports the number of VQE iterations required for selected hydrogen chains when using pre-optimized orbitals and initializing all variational parameters to zero, together with the number of orbital optimization (CASSCF macro-) iterations.
VQE converges within only four iterations, demonstrating that SPA is straightforward to optimize and does not exhibit the optimization difficulties often encountered in generic VQE approaches. 
The orbital optimization requires only five to six macro-iterations, indicating rapid convergence also at this level.
Consequently, the increase in computation time observed in Figure~\ref{fig:times_vs_n} originates primarily from the cost of evaluating a single optimization step rather than from an increase in the number of optimization iterations.\\

\begin{figure}[h]
    \centering
    \includegraphics[width=\columnwidth]{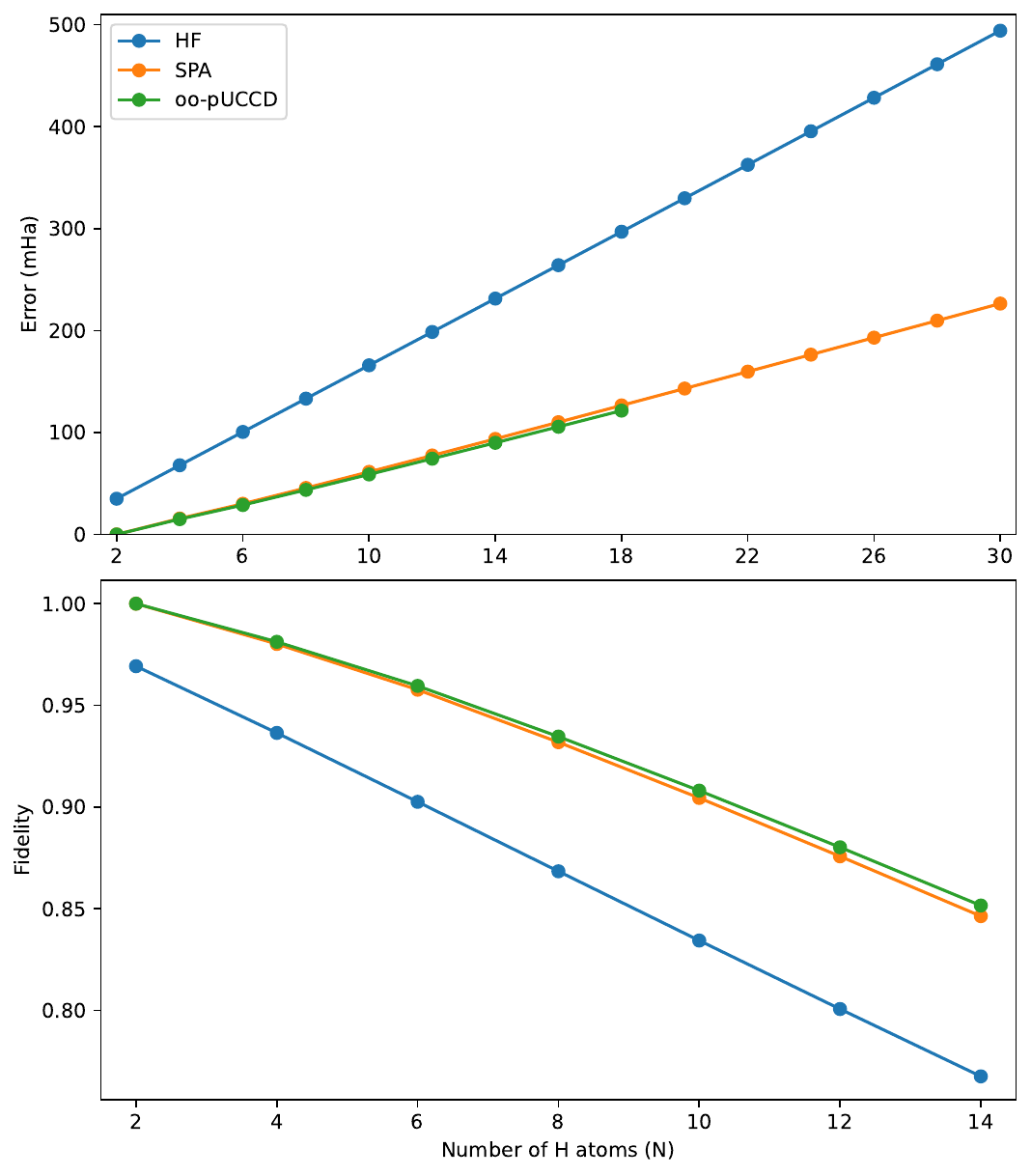}
    \caption{Accuracy of HF, SPA, and oo-pUCCD for linear hydrogen chains at a fixed interatomic distance of $R=1.0~\text{\AA}$. Top: energy errors relative to DMRG. Bottom: Fidelities with respect to FCI.}
    \label{fig:accuracy_vs_n}
\end{figure}

\rev{To assess how the accuracy changes with system size, Figure~\ref{fig:accuracy_vs_n} shows the energy errors (top) and fidelities (bottom) of HF, SPA, and oo-pUCCD. Since FCI becomes computationally impractical for the longer chains, DMRG energies are used as the energy reference up to $\mathrm{H_{30}}$. The oo-pUCCD energy calculations are reported up to $\mathrm{H_{18}}$ because of their greater classical simulation and optimization cost.}

\rev{The HF and SPA errors both increase approximately linearly with system size, but the SPA error remains substantially smaller throughout the investigated range. The additional pair excitations included in pUCCD consistently lower the energy relative to SPA, although the improvement is small for these hydrogen chains. At $\mathrm{H_{18}}$, oo-pUCCD lowers the energy by 5 mHa relative to SPA.
The corresponding fidelities relative to FCI are given in the bottom panel and show the same behavior. SPA gives substantially higher fidelities than HF, while oo-pUCCD provides a smaller additional improvement over SPA. At $\mathrm{H_{14}}$, the fidelity increases from 0.846 for SPA to 0.851 for oo-pUCCD. 
Thus, SPA provides results close to those of oo-pUCCD for these systems while retaining its simpler separable structure and lower classical computational cost.}

\subsection{Molecular benchmarks}

We next assess the extension of SPA to molecular systems, including alkanes up to $\mathrm{C_8H_{18}}$, a diverse set of small molecules ($\mathrm{H_2O}$, $\mathrm{NH_3}$, $\mathrm{CH_4}$, $\mathrm{HF}$, $\mathrm{H_2CO}$, $\mathrm{C_2H_4}$, $\mathrm{C_2H_2}$, and $\mathrm{CH_3OH}$) and conjugated polyenes up to $\mathrm{C_{14}H_{16}}$, at equilibrium geometries in the STO-3G basis.

\subsubsection{Alkanes}

\begin{figure}[h]
    \centering
    \includegraphics[width=\columnwidth]{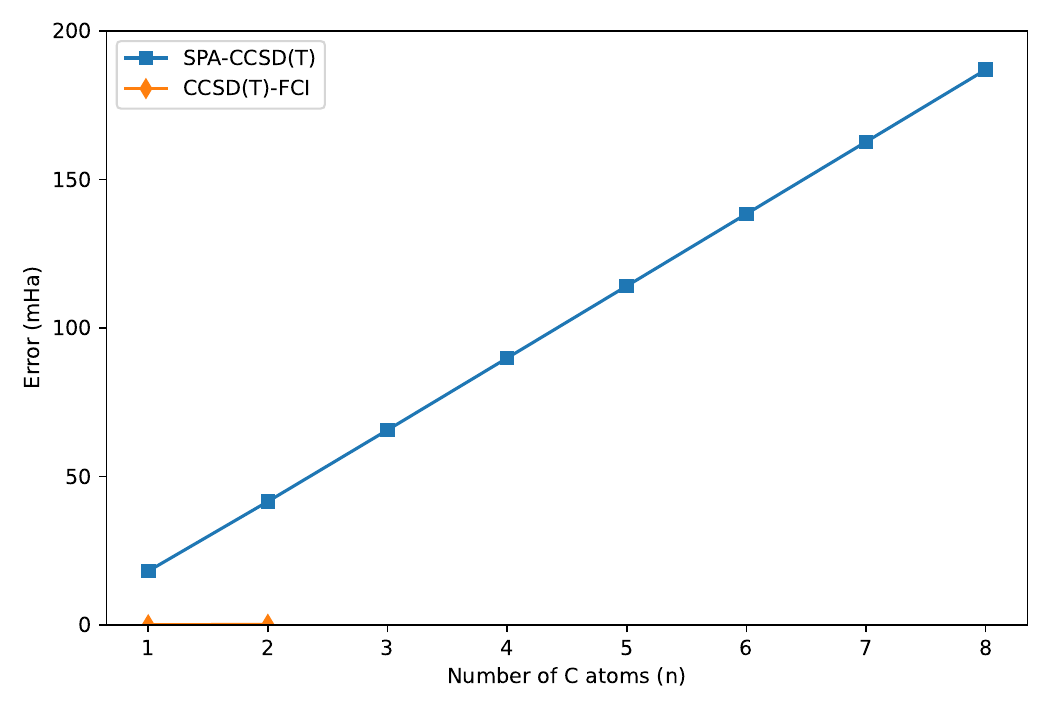}
    \caption{SPA energy errors relative to CCSD(T) for alkanes $\mathrm{C_nH_{2n+2}}$. For $\mathrm{CH_4}$ and $\mathrm{C_2H_6}$, CCSD(T) errors relative to FCI are also shown and found to be negligible, supporting its use as a reference for larger systems. SPA circuits have depth $3$ independent of system size. } 
    \label{fig:alkanes}
\end{figure}

Figure \ref{fig:alkanes} shows the SPA energy error relative to CCSD(T) for the alkane series. 
The error increases approximately linearly with molecular size, analogous to that observed previously for the hydrogen chains. 
For $\mathrm{CH_4}$ and $\mathrm{C_2H_6}$, CCSD(T) is in excellent agreement with FCI, with negligible energy differences, justifying its use as a reference for larger alkanes, for which FCI becomes prohibitively expensive.
For these small molecules, SPA fidelities are 0.991 and 0.978, respectively, indicating very good agreement with the exact wavefunction.

\subsubsection{Small molecules}

\begin{figure}[h]
    \centering
    \includegraphics[width=\columnwidth]{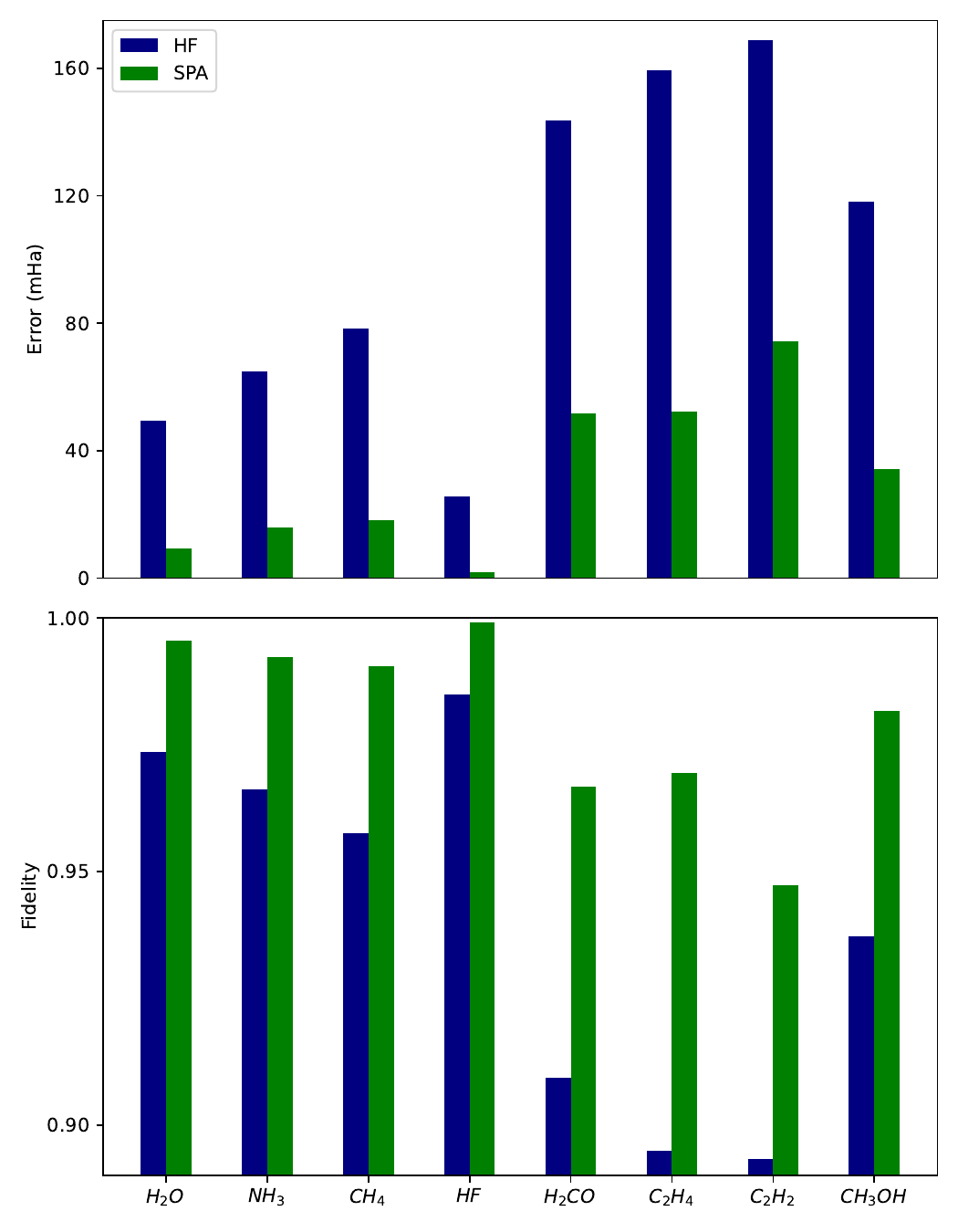}
    \caption{(top) SPA and HF energy errors relative to FCI and (bottom) fidelities $|\braket{\Psi|\Psi_\text{FCI}}|^2$, of a set of small molecules.}
    \label{fig:small-molecs}
\end{figure}

The top panel of Figure \ref{fig:small-molecs} compares SPA and HF energy errors relative to FCI for a set of small molecules.
SPA consistently improves upon HF across all systems, although never reaching chemical accuracy (1.6 mHa) except for hydrogen fluoride. 
For molecules containing only single bonds, the SPA energy error remains below $35$ mHa, while systems with double and triple bonds exhibit larger deviations of approximately $50$ mHa and $75$ mHa, respectively.
The bottom panel of Figure \ref{fig:small-molecs} shows the corresponding wavefunction fidelities with respect to FCI. While HF already yields high fidelities for these systems, SPA is systematically closer to the FCI solution, reaching near-perfect agreement in weakly correlated cases.

\subsubsection{Conjugated $\pi$-systems}

\begin{figure}[h]
    \centering
    \includegraphics[width=\columnwidth]{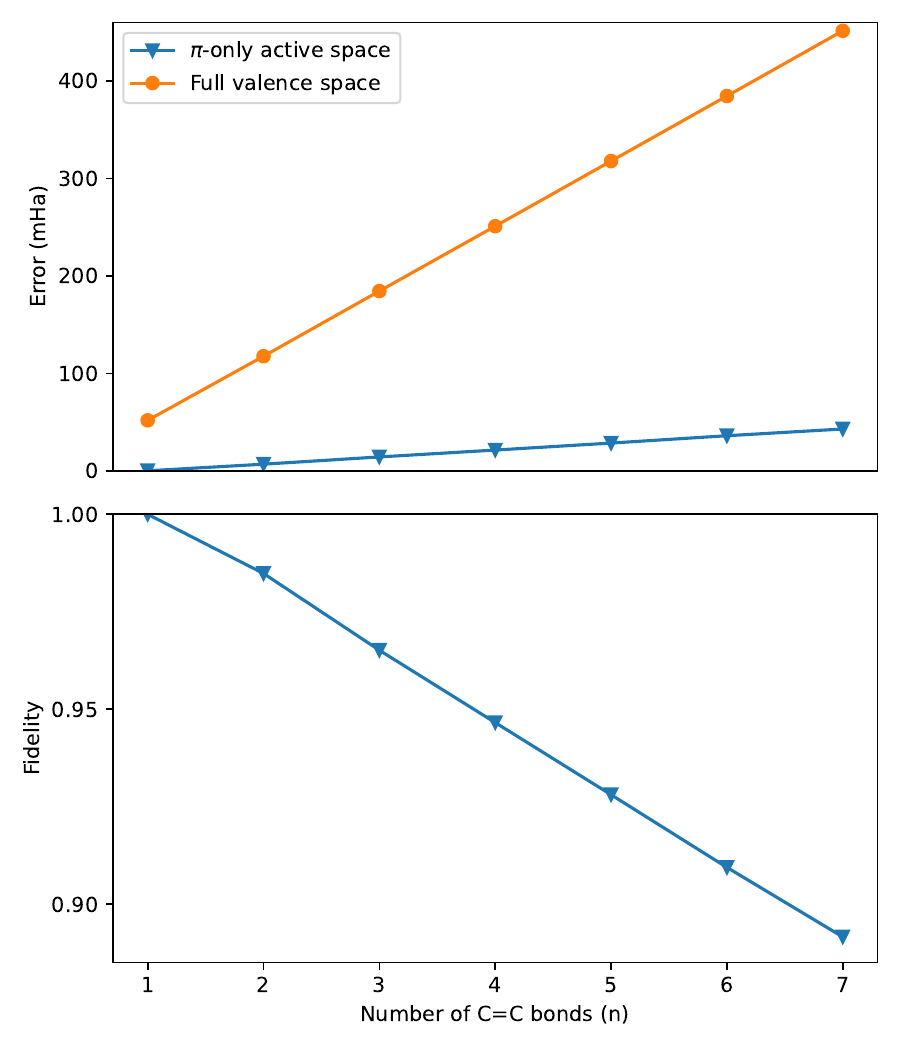}
    \caption{SPA results for the conjugated polyene series $\mathrm{C_{2n}H_{2n+2}}$. Top: SPA energy error in the \rev{full valence orbital space} relative to CCSD(T) (orange), and in the $\pi$-only active space relative to FCI (blue). Bottom: corresponding wavefunction fidelity in the $\pi$ space with respect to FCI.} 
    \label{fig:polyenes}
\end{figure}

To assess SPA in the presence of delocalized electrons, we consider the conjugated polyene series $\mathrm{C_{2n}H_{2n+2}}$ up to $n=7$.
Figure~\ref{fig:polyenes} (top) shows the SPA energy error relative to CCSD(T) in the rev{full valence orbital space}.
To isolate the role of the delocalized $\pi$ electrons, we further performed calculations in active spaces containing only the $\pi$ orbitals, for which FCI remains tractable. The corresponding SPA energy error relative to FCI is also shown in the top panel of Figure~\ref{fig:polyenes}, while the wavefunction fidelity is shown in the bottom panel.

In all cases, the error increases linearly with system size. 
In the \rev{full valence orbital space}, the SPA energy error increases from approximately $50$ to $450$ mHa, while in the $\pi$-only subspace it remains below $45$ mHa across the entire series, with the corresponding fidelity remaining above $0.89$.
The significantly lower error in the $\pi$-only calculations indicates that the $\pi$ subsystem is well described within SPA, and that the dominant contribution to the larger \rev{full valence-space} error does not originate from it. All in all, the behaviour is consistent with the hydrogenic model systems and the linear chain of alkanes, where the error in the $\pi$-system can be attributed to the missing resonance structures while the error in the \rev{full valence-space} ($\pi$ and $\sigma$) is the linearly accumulating error from the separable form of the state. 

\section{Conclusion and Outlook}

In summary, we have benchmarked the SPA across linear hydrogen chains, alkanes, a set of small molecules and conjugated polyenes.
The results show that SPA provides a stable and qualitatively correct description of ground-state energies, in particular in strongly correlated regimes where classical single-reference methods such as HF and CCSD(T) break down.
At the same time, the ansatz exhibits favourable computational scaling, with a constant circuit depth and a number of variational parameters that grows linearly with system size.
\rev{However, systematically attaining high quantitative accuracy is limited by the restriction to the seniority-zero subspace, which excludes configurations involving broken electron pairs. The oo-pUCCD comparison presented here indicates that removing the additional separability restriction while remaining within the seniority-zero space yields only modest improvements for the hydrogen chains considered. SPA may therefore be useful as a standalone method when qualitative accuracy is sufficient, while its main role is to provide a compact and chemically informed initial state for subsequent treatments that recover correlation outside the paired-electron space.}

A natural direction for future work is therefore the incorporation of post-SPA correlation corrections.
In particular, recent work \cite{zhao2023orbital} on the pair-based ansatz oo-upCCD demonstrated that adding a second-order perturbative correction~\cite{zhao2024enhancing} significantly reduces energy errors, while remaining compatible with quantum hardware and robust against noise.
Adapting such a scheme to the SPA framework could improve quantitative accuracy while preserving its favourable scaling and hardware efficiency.\\
The methodology is classically tractable and results in extremely shallow quantum circuits. Compared to the established HF method, that comes with comparable computational cost, the method consistently gives better energies and fidelities while most-importantly behaves more consistent in regimes where the HF approximation breaks.\\
Currently, the computational bottleneck of our implementation is the orbital optimization that interfaces the standard orbital optimizer of \texttt{pyscf}. In the future, this can be mitigated by using orbital optimizers that exploit hardcore bosonic structures~\cite{limacher2026attacking}. Recent works in machine learning also demonstrated successful transfer learning~\cite{bincoletto2026transfer} on angles for SPA circuits and potential extension to orbital coefficients~\cite{vanderhorst2026transferablemachinelearningapproach}.

\section*{Data Availability}
The scripts used to generate the data, together with the raw benchmark datasets (CSV files) and instructions for reproducing the results, are available at \\ \url{https://github.com/lily-barta/spa-benchmark}.

\section*{Conflicts of interest}
There are no conflicts of interest to declare.

\section*{Acknowledgement}
The authors acknowledge support from the Federal Ministry of Research, Technology and Space (BMFTR) of Germany through the VeriVaQ project (13N17145).
JSK gratefully acknowledges support from the Hightech Agenda Bayern. We thank Francisco Javier Del Arco Santos, Jonas Motyl and Viktoria Raba for various discussions. Computational resources were in-part provided by the LiCCA HPC cluster of the University of Augsburg, co-funded by the German Research Foundation (DFG) – Project-IDs 499211671 \& 572310035.

%% file: appendix.tex
\rev{\section{Sensitivity to orbital localization and ambiguous initial pairing}\label{app:orb-loc}}

\rev{We assess the sensitivity of SPA to orbital localization and initial pairing using butane and 1,3-butadiene in the STO-3G basis with frozen core orbitals. 
We compare CLPO, Boys, and Pipek–Mezey (PM) localizations. 
For Boys and PM orbitals, the orbital pairing was selected manually. Where the localization did not yield orbitals clearly associated with individual bonds, several plausible alternative pairings were tested using the same localized orbitals. This comparison samples the pairing ambiguity without exhaustively enumerating all possible assignments. Each electron-pair subset contains two spatial orbitals.}

\rev{Tables~\ref{tab:init_butadiene} and~\ref{tab:init_butane}
report the SPA energies before and after orbital optimization, together with the number of orbital-optimization macroiterations. Here, $E_{\mathrm{loc}}$ denotes the SPA energy in the initial localized orbital basis, and $E_{\mathrm{opt}}$ denotes the energy after orbital optimization. The labels B1--B3 and P1--P3 distinguish the tested Boys and PM pairings, respectively.
}

\begin{table}[h]
\centering
\setlength{\tabcolsep}{8pt}
\caption{Sensitivity to orbital localization and initial pairing for 1,3-butadiene in STO-3G with frozen core orbitals. Energies are given in Hartree. $N_{\mathrm{macro}}$ is the number of orbital-optimization macroiterations.}
\label{tab:init_butadiene}
\begin{tabular}{lccc}
\hline
Initialization
& $E_{\mathrm{loc}}$
& $E_{\mathrm{opt}}$
& $N_{\mathrm{macro}}$ \\
\hline
CLPO     & -153.0964 & -153.2077 &  5 \\
Boys B1  & -153.1488 & -153.2077 & 18 \\
Boys B2  & -153.1419 & -153.2077 & 18 \\
Boys B3  & -153.1437 & -153.2077 & 18 \\
PM       & -153.2059 & -153.2077 &  6 \\
\hline
\end{tabular}
\end{table}

\begin{table}[h]
\centering
\setlength{\tabcolsep}{8pt}
\caption{Sensitivity to orbital localization and initial pairing for butane in STO-3G with frozen core orbitals.
Energies are given in Hartree. $N_{\mathrm{macro}}$ is the number of orbital-optimization macroiterations.}
\label{tab:init_butane}
\begin{tabular}{lccc}
\hline
Initialization
& $E_{\mathrm{loc}}$
& $E_{\mathrm{opt}}$
& $N_{\mathrm{macro}}$ \\
\hline
CLPO     & -155.5730 & -155.6554 &  4 \\
Boys B1  & -155.5495 & -155.6554 & 10 \\
Boys B2  & -155.5495 & -155.6554 & 10 \\
PM P1    & -155.5467 & -155.6554 & 13 \\
PM P2    & -155.5465 & -155.6554 & 12 \\
PM P3    & -155.5465 & -155.6554 & 13 \\
\hline
\end{tabular}
\end{table}

\rev{Across the tested localizations and two-orbital pairings, the energy spread decreases from approximately 110 mHa before orbital optimization to 1 $\mu$Ha afterward for butadiene, and from 26 mHa to 1 $\mu$Ha for butane.
Thus, substantial differences in the initial energies are strongly reduced by orbital optimization.
For a fixed localized orbital basis, the different plausible pairings converge to essentially the same optimized energy.
The remaining differences are on the microhartree scale.
Thus, orbital optimization largely removes the dependence on the initial pairing for the alternatives tested here.}

\rev{The initialization nevertheless affects the convergence effort. CLPO requires 5 macroiterations for butadiene and 4 for butane, whereas the other initializations require 6--18 and 10--13, respectively. In these examples, CLPO therefore provides an efficient starting point, while orbital optimization strongly mitigates the energy sensitivity to both localization and initial pairing. These observations concern the tested geometries and two-orbital partitions and do not establish initialization independence nor guaranteed convergence for arbitrary molecular systems.}

\clearpage
\rev{\section{Automated hydrogen calculations and pairing heuristics benchmarks}
\label{app:hydrogen}}

\rev{The \textsc{quanti-gin} package implements the geometry-based pairing procedure described in Section \ref{sec:algorithm} and the subsequent SPA optimization for hydrogen systems. It supports linear chains, rings, and random hydrogen configurations. The following minimal example illustrates its use for a specified $\mathrm{H}_4$ geometry in the STO-3G basis. The initial pairing is determined automatically using the scaled-blossom heuristic, without orbital localization or a manually supplied pair assignment. An FCI calculation is included as a reference for this small example.}

\begin{lstlisting}
import tequila as tq
from quanti_gin.data_generator import DataGenerator
from quanti_gin.shared import blossom_scaled

geometry = """
H 0.0 0.0 0.0
H 0.0 0.0 1.5
H 0.0 0.0 2.24
H 0.0 0.0 3.74
"""
coordinates = DataGenerator.parse_geometry_string(geometry)

molecule = tq.Molecule(
    geometry=geometry,
    basis_set="sto-3g",
    units="angstrom",
    backend="pyscf",
)

edges = blossom_scaled(len(coordinates), coordinates)
result = DataGenerator.run_spa_optimization(
    molecule=molecule,
    coordinates=coordinates,
    edges=edges,
)

spa_energy = result["energy"]
fci_energy = molecule.compute_energy("fci")

print(f"Edges: {edges}")
print(f"SPA: {spa_energy:.10f} Eh")
print(f"FCI: {fci_energy:.10f} Eh")
print(
    f"SPA - FCI: "
    f"{1000 * (spa_energy - fci_energy):.6f} mEh"
)
\end{lstlisting}

\rev{Alternatively, a random geometry can be generated by replacing the geometry definition and parsing step with:}

\begin{lstlisting}
coordinates = DataGenerator.generate_coordinates(
    count=4, max_distance=2
)
geometry = DataGenerator.generate_geometry_string(coordinates)
\end{lstlisting}

\begin{figure}[h]
\centering
\includegraphics[width=\columnwidth]{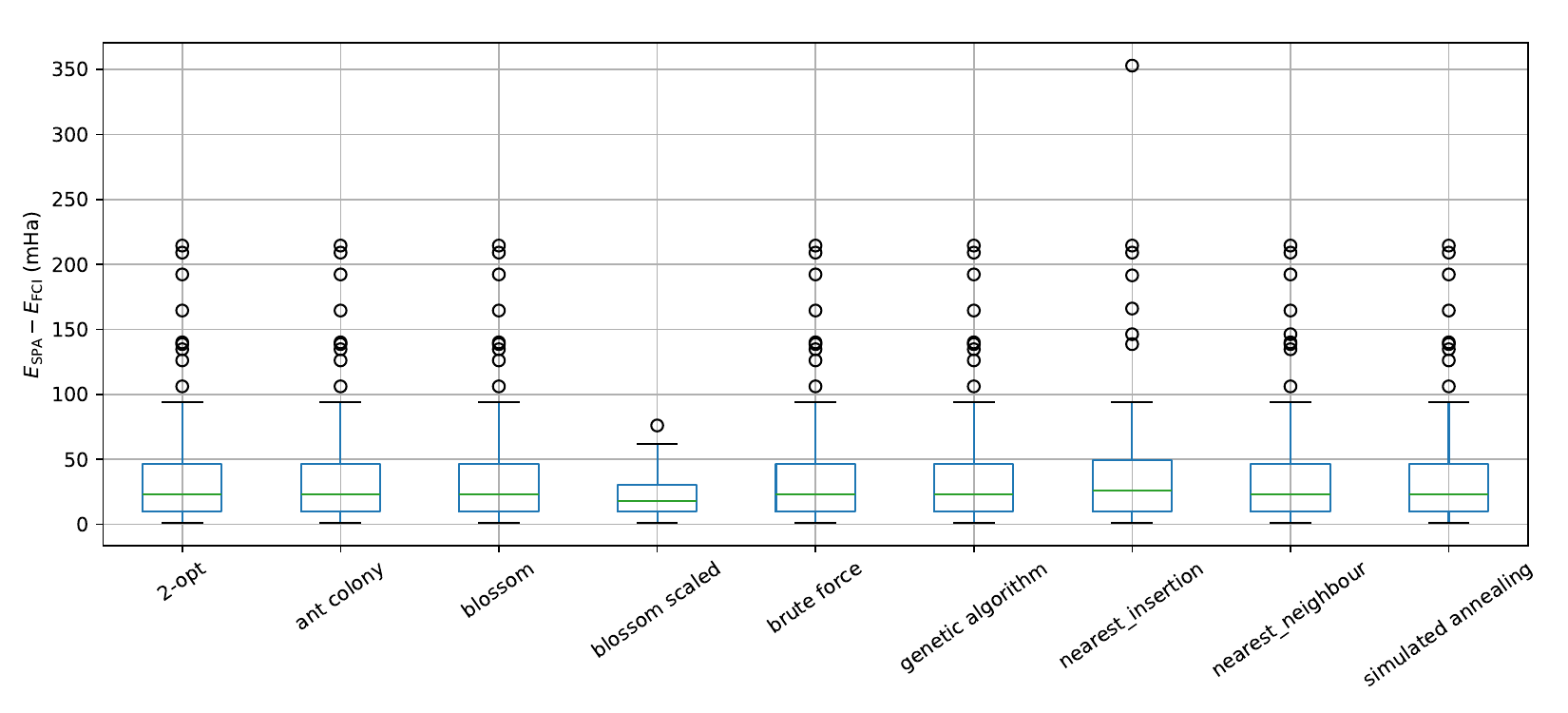}
\caption{
Distributions of orbital-optimized SPA energy errors
relative to FCI for nine initial pairing heuristics,
evaluated on 100 random $\mathrm{H}_6$ geometries in STO-3G. For more information see the online resource from Stein/Raba.~\cite{stein-raba}
}
\label{fig:h6_pairing}
\end{figure}

\rev{To compare the automated pairing prescriptions, we evaluate
the nine heuristics implemented in \textsc{quanti-gin} on 100 random $\mathrm{H}_6$ geometries in STO-3G.
Figure~\ref{fig:h6_pairing} shows the distributions of
orbital-optimized SPA energy errors relative to FCI.
The benchmark script is available in the accompanying
code repository \cite{spa-repo}.}

\rev{The scaled-blossom heuristic yields a lower median error, a narrower interquartile range, and a smaller large-error tail than the other heuristics in this sample.
These results support its use as the default pairing prescription for the automated hydrogen workflow.}

\rev{This comparison addresses the ability of the heuristics to select suitable initial pairings from the nuclear geometry.}
\rev{It therefore complements Appendix \ref{app:orb-loc}, which examines alternative plausible pairings within a fixed localized orbital basis. The agreement obtained between those plausible alternatives does not imply that orbital optimization can compensate for every initial assignment.}

